\documentclass[twocolumn,showpacs,preprintnumbers,amsmath,amssymb,pra,aps,superscriptaddress,floatfix]{revtex4-1}

\usepackage{graphicx}  
\usepackage{latexsym}
\usepackage{physics}
\usepackage{color}
\usepackage{hyperref}
\hypersetup{
    colorlinks=true
}

\graphicspath{{../figures/}}

\begin{document}

\title{Rapid high-fidelity gate-based spin read-out in silicon}

\author{G.~Zheng}
\affiliation{QuTech and Kavli Institute of Nanoscience, Delft University of Technology, Lorentzweg 1, 2628 CJ Delft, The Netherlands}
\author{N.~Samkharadze}
\affiliation{QuTech and Kavli Institute of Nanoscience, Delft University of Technology, Lorentzweg 1, 2628 CJ Delft, The Netherlands}
\author{M.~L.~Noordam}
\affiliation{QuTech and Kavli Institute of Nanoscience, Delft University of Technology, Lorentzweg 1, 2628 CJ Delft, The Netherlands}
\author{N.~Kalhor}
\affiliation{QuTech and Kavli Institute of Nanoscience, Delft University of Technology, Lorentzweg 1, 2628 CJ Delft, The Netherlands}
\author{D.~Brousse}
\affiliation{QuTech and Netherlands Organization for Applied Scientific Research (TNO), Stieltjesweg 1 2628 CK Delft, The Netherlands}
\author{A.~Sammak}
\affiliation{QuTech and Netherlands Organization for Applied Scientific Research (TNO), Stieltjesweg 1 2628 CK Delft, The Netherlands}
\author{G.~Scappucci}
\affiliation{QuTech and Kavli Institute of Nanoscience, Delft University of Technology, Lorentzweg 1, 2628 CJ Delft, The Netherlands}
\author{L.~M.~K.~Vandersypen}
\affiliation{QuTech and Kavli Institute of Nanoscience, Delft University of Technology, Lorentzweg 1, 2628 CJ Delft, The Netherlands}

\date{\today}

\begin{abstract}
Silicon spin qubits form one of the leading platforms for quantum computation~\cite{Zwanenburg13,Vandersypen17}. As with any qubit implementation, a crucial requirement is the ability to measure individual quantum states rapidly and with high fidelity. As the signal from a single electron spin is minute, different spin states are converted to different charge states~\cite{Ono02,Elzerman04}. Charge detection so far mostly relied on external electrometers~\cite{Hanson07,Field93,Barthel10}, which hinders scaling to two-dimensional spin qubit arrays~\cite{Vandersypen17,Veldhorst17,Li18}. As an alternative,  gate-based dispersive read-out based on off-chip lumped element resonators were introduced~\cite{Colless13,Zalba15,Betz15,House15,Rossi17,Ahmed18}, but here integration times of 0.2 to 2 ms were required to achieve single-shot read-out~\cite{Pakkiam18,West18,Urdampilleta18}.  Here we connect an on-chip superconducting resonant circuit to two of the gates that confine electrons in a double quantum dot. Measurement of the power transmitted through a feedline coupled to the resonator probes the charge susceptibility, distinguishing whether or not an electron can oscillate between the dots in response to the probe power. With this approach, we achieve a signal-to-noise ratio (SNR) of about six within an integration time of only 1 $\mu$s. Using Pauli's exclusion principle for spin-to-charge conversion, we demonstrate single-shot read-out of a two-electron spin state with an average fidelity of $>$98\% in 6 $\mu$s. This result may form the basis of frequency multiplexed read-out in dense spin qubit systems without external electrometers, therefore simplifying the system architecture.
\end{abstract}

\maketitle

Single-shot read-out is required for implementing quantum error correcting schemes~\cite{Fowler12}, where the read-out and correction should be performed with high fidelity and well within the qubit coherence times. Spin qubits are commonly measured using spin-to-charge conversion in combination with various types of charge detectors nearby the qubit dots ~\cite{Hanson07}. Among those, an ancillary quantum dot probed using radio-frequency reflectometry (RF-dot) is the most sensitive charge detector~\cite{Barthel10}. However, dense spin qubit architectures don't allow space for integrating  detectors adjacent to the qubit dots~\cite{Vandersypen17,Veldhorst17,Li18}. Therfore, applying RF-reflectometry to one or more gates that are already in place to define the qubit dots, a technique known as gate-based dispersive read-out, has been an ongoing research topic across different semiconductor platforms~\cite{Colless13,Zalba15,Betz15,House15,Rossi17,Ahmed18}. However, so far the tank circuits used a commercial or superconducting inductor mounted on a printed circuit board adjacent to the quantum dot chip. These circuits are quite lossy and contain a large parasitic capacitance, masking the useful signal from the capacitive response of the quantum dots. Even though single-shot read-out of spin states could be achieved thanks to long spin relaxation timescales, the effective detection bandwidths were limited by the SNR to a few kilohertz~\cite{Pakkiam18,West18,Urdampilleta18}. 

Here, we use a fully integrated on-chip superconducting resonator in the GHz-range to perform single-shot singlet-triplet read-out. The high quality factor and large impedance of the resonator enable fast high-fidelity charge detection. The resonator linewidth of $\sim$2.2 MHz sets the maximum measurement bandwidth, and we obtain a SNR of 6 at a $\sim$350 kHz bandwidth. Conventional Pauli spin blockade was used to map the two-electron spin state onto the charge state. Despite a comparatively short $T_1 \approx 160$ $\mu$s, the achieved single-shot spin read-out fidelity was $>98\%$. 

\begin{figure}
	\includegraphics[width=\columnwidth]{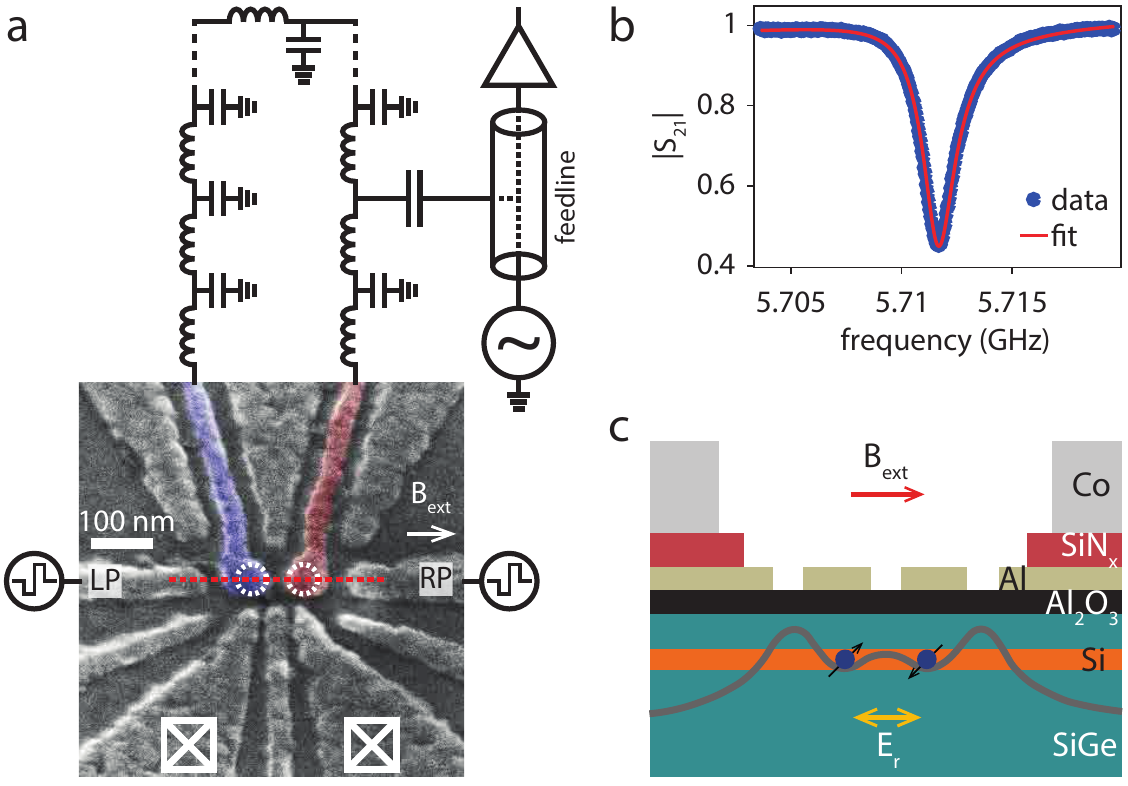}
	\caption{\label{f1}
		\textbf{Device schematics.}
		\textbf{a} Scanning electron micrograph of a device nominally identical to that used in the experiment, showing the Al gate electrodes for accumulation and confinement of electrons, with a schematic of the superconducting resonanator. The circuit consists of a high impedance NbTiN thin wire with a $\lambda/2$ resonance mode. The ends of this wire, the red and purple shaded gates, extend towards and overlap with the location of the two dots (white dashed circles). The LP and RP gates are used to adjust the electrochemical potentials of the dots. Voltage pulses are sent to these gates as well, through bias tees. The white crossed boxes at the bottom indicate the location of Fermi reservoirs of electrons that are connected to source and drain electrodes outside the image.
		\textbf{b} Normalized transmission amplitude through the superconducting co-planar transmission line (feedline), prior to the formation of dots. From a Lorentzian fit (red line) the resonance frequency $f_0=5.7116$ GHz, loaded quality factor $Q\approx2600$, internal quality factor $Q_i\approx5780$ and coupling quality factor $Q_c\approx4730$ are extracted~\cite{Khalil12}.
		\textbf{c} Schematic cross-section of the device along the red dashed line in \textbf{a}. A two-dimensional electron gas is formed in the $\sim$10 nm thick Si strained quantum well by positively biasing the Al gates. The SiGe barrier on top of the quantum well is approximately 30 nm thick. The resonator gates produce a tiny oscillating electric field $E_{\mathrm{r}}$ to which the electron in the DQD responds. Co micromagnets are located on top of the gate stack, isolated from the gates by a layer of SiN dielectric, and provide a transverse field gradient after they are magnetized by an external magnetic field $B_{\mathrm{ext}}$. This gradient is not used intentionally in this experiment, but may impact the spin relaxation time $T_1$.
	}
\end{figure}

A top view of a Si/SiGe double quantum dot (DQD) device is depicted in Fig.~1a~\cite{Samkharadze18}. The DQD confining the electrons is formed in the strained Si quantum well of a Si/SiGe heterostructure by applying appropriate voltages to a single layer of metallic gates to create a double-well potential~\cite{Rochette17} (see Fig.~1c). The device is cooled down to $\sim$20 mK in a dilution refrigerator. The colored gates are galvanically connected to the NbTiN superconducting nanowire resonator~\cite{Samkharadze16}, which can be modeled as a distributed network with a characteristic impedance of $\sim$1 k$\Omega$. This resonator is capacitively coupled to a planar superconducting transmission line (feedline) through which a microwave signal with a power  $P = -110$ dBm is sent to probe the resonant circuit. At this power $\sim$1 photon is stored in the resonator. We measure the transmission amplitude and phase near 5.71 GHz in a standard heterodyne scheme after amplification at 4 K and room temperature. The observed dip in the normalized transmission amplitude in Fig.~1b reveals the resonance frequency $f_0=5.7116$ GHz, along with the total linewidth $\kappa/2\pi\approx 2.2$ MHz. This corresponds to a loaded quality factor $Q=f_0/(\kappa/2\pi)\approx2600$.

\begin{figure}
	\includegraphics[width=\columnwidth]{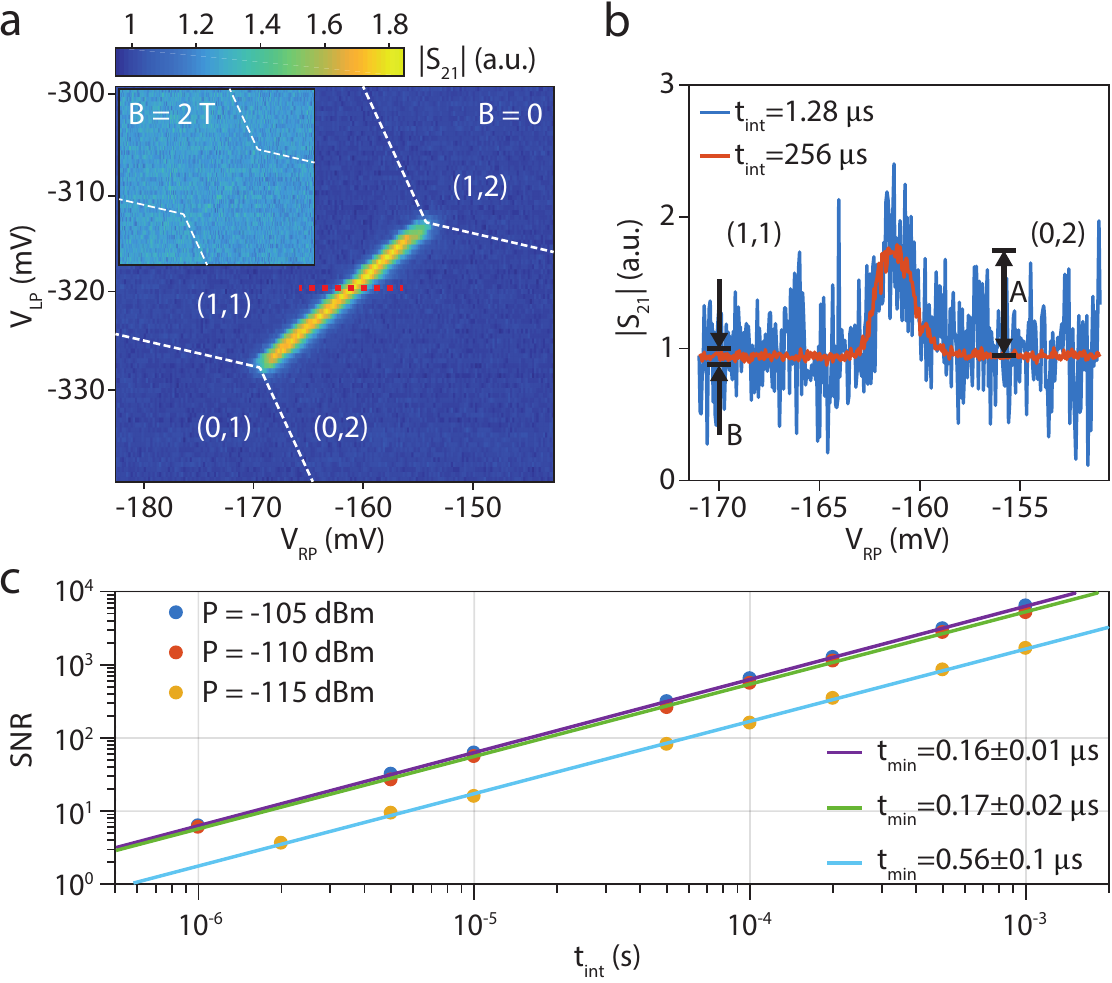}
	\caption{\label{f2}
		\textbf{Characterization of the charge sensitivity.}
		\textbf{a} The transmitted amplitude at 5.7116 GHz as a function of the voltages applied to LP and RP. The yellow bright line defines the zero detuning axis, along which one electron can tunnel freely between the left and right dots while a second electron remains in the right dot. The white dashed lines indicate the boundaries of the charge stability diagrams, where electrons can be added to or removed from the dots. The scan was taken by repeatedly applying a 500 Hz sawtooth wave to RP and stepping $V_{\mathrm{LP}}$ every 200 ms. Each pixel in the plot has an effective integration time of 1 ms. The inset shows the exact same scan in the presence of an external in-plane magnetic field of 2 T. Here the probe frequency was set to 5.6930 GHz, to account for a shift of the resonance frequency with magnetic field. The overall transmission in the new frequency range is higher. The white dashed lines in the inset are copied from the main plot.
		\textbf{b} The transmitted amplitude versus the voltage on RP around zero detuning (red dashed line in \textbf{a}). Data is collected point by point in $V_{\mathrm{RP}}$, with an integration time of 1.28 $\mu$s (blue trace) and 256 $\mu$s (red trace).
		\textbf{c} SNR as a function of the integration time. Three sets of data are shown, corresponding to a power of -105 dBm (blue dots), -110 dBm (red dots) and -115 dBm (yellow dots) through the feedline. The red data points were taken in a slightly different charge configuration from the blue and yellow data points. Each data set is fit well by a straight line (solid lines), from which we extrapolate $t_{\mathrm{min}}$, the integration time for SNR = 1. The rms noise amplitude $B$ was obtained from time traces containing 1000 points for each integration time. The errors in $A$ and $B$ translate to uncertainties (standard deviations) in SNR that are smaller than the size of the data points.
	}
\end{figure}

The resonator is a sensitive probe that can detect tiny changes in the charge susceptibility of the DQD~\cite{Cottet11,Petersson12,Schroer12,Frey12,Chorley12}. This susceptibility is largest at zero detuning, $\varepsilon = 0$, where the electrochemical potentials of the left and right dots align and an electron is able to tunnel freely between the two dots. In this case, the DQD damps the resonator and shifts its frequency~\cite{Mi17,Stockklauser17,Bruhat18}. Away from zero detuning, the electron(s) can only move within a quantum dot, and the electrical susceptibility is negligible in comparison. By recording the transmitted signal at the resonance frequency $f_0$ while varying the voltages on the plunger gates, LP and RP, one can map out the charge stability diagram of the DQD. A typical diagram in the few-electron regime is shown in Fig.~2a, where $(N_{L},N_{R})$ indicates the charge occupation, with $N_{L}$ $(N_{R})$ the number of electrons in the left (right) dot. A bright yellow line appears at the transition between the (1,1) and (0,2) charge states. Since the probe frequency of around 5.7 GHz is above the interdot tunnel coupling $t_c\approx2$ GHz, measured using two-tone spectroscopy~\cite{Schuster05}, the system is not in the adiabatic limit where quantum capacitance arising from the curvature of the dispersion relation dominates the response~\cite{Mizuta17}. Instead, there is also a significant contribution from the tunneling capacitance, whereby charges non-adiabatically redistribute in the double dot at a rate comparable to the probe frequency.

We first quantify the sensitivity of the resonator to changes in the DQD susceptibility due to electron tunneling. We scan over the interdot transition by sweeping the voltage on RP (red dashed line in Fig.~2a). Fig.~2b shows two examples of the resulting line traces, with an integration time of 1.28 $\mu$s (blue) and 256 $\mu$s (red) per point. The power SNR is defined as $\mathrm{SNR}=(A/B)^2$. The signal $A$ is the difference between the transmitted amplitude at the interdot transition ($V_{\mathrm{RP}}\approx-162$ mV) and the amplitude in the Coulomb blockaded region, where no electrons are allowed to tunnel. This difference is obtained from a Gaussian fit to data such as that in Fig.~2b. The noise $B$ is the root-mean-square (rms) noise amplitude measured with the electrons in Coulomb blockade ($V_{\mathrm{RP}}\approx-170$ mV). We expect $A^2$ to increase linearly with the probe power, and $B^2$ to decrease linearly with the integration time. Fig.~2c shows the $\mathrm{SNR}$ as a function of the integration time for three different probe powers. The data points follow $\mathrm{SNR}(t_{\mathrm{int}})=t_{\mathrm{int}}/t_{\mathrm{min}}$, with $t_{\mathrm{min}}$ the integration time corresponding to an $\mathrm{SNR}$ of unity. We find $t_{\mathrm{min}} \approx$ 170 ns at -110 dBm input power, and it is $\sim$3.3 times longer than at -115 dBm, which is expected from the 5 dB difference in power. At higher power (-105 dBm) $t_{\mathrm{min}}$ begins to saturate, presumably since the electron displacement in the DQD reaches a maximum. The inverse resonator linewidth imposes an additional constraint on the measurement time of $0.35(\kappa/2\pi)^{-1} \approx$ 160 ns. Using the standard definition of the charge sensitivity, we get $\delta q=e\sqrt{t_{\mathrm{min}}}=(4.1\pm0.3) \times 10^{-4}$  $e/\sqrt{\mathrm{Hz}}$ at $P=-110$ dBm. This is an order of magnitude higher than reported for a microwave resonator probed with a quantum-limited Josephson parametric amplifier (JPA) but two orders of magnitude lower compared to the value reported without JPA~\cite{Stehlik15}.

\begin{figure}
	\includegraphics[width=\columnwidth]{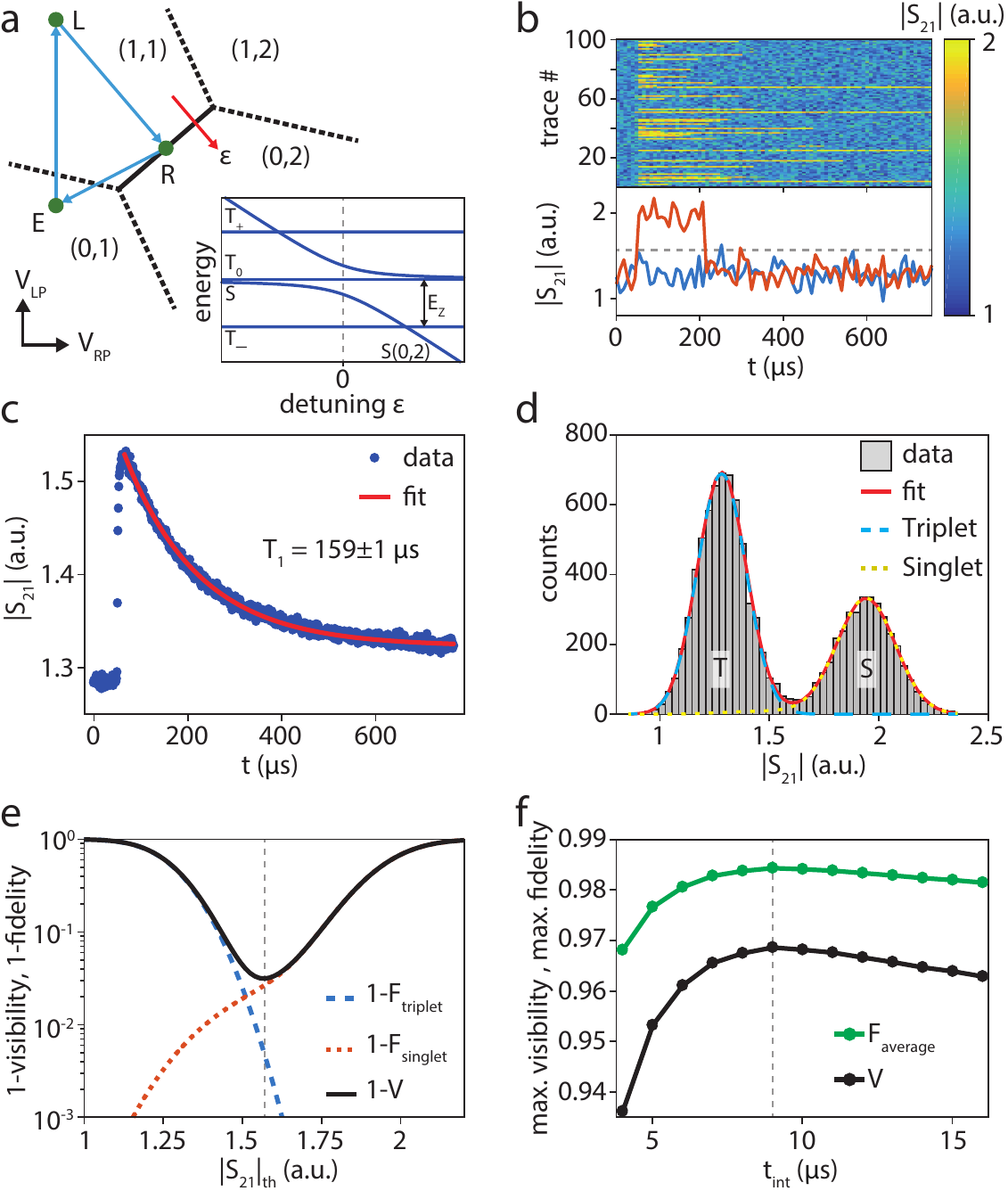}
	\caption{\label{f3}
		\textbf{Single-shot spin read-out and fidelity analysis.}
		\textbf{a} Schematic of a typical charge stability diagram with a three-stage pulse sequence. The DC voltages on LP and RP are set to point R. Voltage pulses are applied to LP and RP as well. A two-electron energy level diagram as a function of the detuning $\varepsilon$ (red arrow) in finite magnetic field is shown in the lower half. The $T_{\pm}$ states are separated from $T_0$ by $E_Z=g\mu_B B$, with $g$ the g-factor and $\mu_B$ the Bohr magneton. Valley-orbit states are neglected in this diagram for simplicity.
		\textbf{b} One hundred single-shot traces with 9 $\mu$s averaging time per data point are shown in the top panel as a function of the measurement time. The traces start 50 $\mu$s prior to pulsing to point R. The bright yellow lines correspond to the signal from the spin singlet state. Two traces are shown separately in the bottom panel. For the blue (red) line the electron was loaded into either T$_0$(1,1) or T$_-$(1,1) (S(1,1)) at point L. The grey dashed line represents the optimum threshold $|S_{21}|_{\mathrm{th}}\approx1.57$.
		\textbf{c} The transmitted amplitude as a function of measurement time shows a typical $T_1$ decay. The data is taken at zero detuning (point R) and 2 T.
		\textbf{d} Measured histogram of the single-shot traces with 9 $\mu$s integration time. A model adapted from Ref.~\cite{Barthel09} was fitted to the data (red solid line) to extract the triplet (blue dashed line) and singlet (yellow dotted line) distributions. See main text for details.
		\textbf{e} The calculated spin read-out fidelities and visibility as a function of threshold amplitude for 9 $\mu$s integration. The maximum visibility is found by setting the threshold at $|S_{21}|_{\mathrm{th}}\approx1.57$.
		\textbf{f} Maximum average fidelity and visibility as a function of the integration time. For each $t_{\mathrm{int}}$ a similar analysis as presented in \textbf{e} was performed.
	}
\end{figure}

Having characterized the charge sensitivity, we now move on to detecting spin states. 
At $\varepsilon=0$, the S(1,1) and S(0,2) singlet states hybridize due to finite interdot tunnel coupling $t_c$. Thus when the system is in a singlet state, one electron is allowed to tunnel between the dots, loading the resonator as a result. 
When the system is in one of the triplet states, there is negligible hybridization of the (1,1) and (0,2) states at $\varepsilon=0$ (the valley splitting is estimated to be $\sim$85 $\mu$eV from magnetospectroscopy), so tunneling is now prohibited and the resonator remains unaffected. 
At zero magnetic field the two electrons form a spin singlet ground state and a strong signal is observed at zero detuning, as discussed (Fig.~2a). When we apply an external in-plane magnetic field $B_{\mathrm{ext}}$ of 2 T, the triplet state T$_-$(1,1) becomes the ground state (see Fig.~3a). As expected, this suppresses the signal from the S(1,1)-S(0,2) tunnelling significantly (see inset Fig.~2a). 

We next probe the spin dynamics of our system by applying voltage pulses to gates LP and RP (see Fig.~3a), first to empty the left quantum dot at point E (100 $\mu$s), then to load an electron with a random spin orientation into the left dot at point L (10 $\mu$s), and finally to measure the response of the resonator at point R. 
We perform 10000 repetitions of this single-shot cycle, and record time traces of the transmitted signal with an integration time of 1 $\mu$s. The traces start 50 $\mu$s prior to pulsing to point R. The results from 100 cycles are shown in Fig.~3b (top panel) with an additional 9 $\mu$s integration time set in post-processing of the experimental data. We perform threshold detection, declaring singlet (triplet) when the signal exceeds (does not exceed) a predefined threshold, $|S_{21}|_{\mathrm{th}}$. Two examples of single traces are shown separately in the bottom panel. The blue trace reflects the case in which the two electrons form a spin triplet state, i.e., the signal remains low during the entire trace. The red trace corresponds to loading a spin singlet state, which here decays to a T\_(1,1) state after $\sim$150 $\mu$s. When averaged over all traces, we obtain a characteristic decay with a relaxation time $T_1$ from the singlet to the triplet ground state of $159 \pm 1$ $\mu$s (see Fig.~3c). Even though this value of $T_1$ is smaller than typical values for silicon devices, we can achieve high-fidelity single-shot read-out thanks to the high sensitivity and bandwidth of our resonator.

In order to characterize the spin read-out fidelity, we create a histogram of the signal integrated over the first 9 $\mu$s in point R. A clear bimodal distribution is visible in Fig.~3d. We fit the data to a model that is based on two noise-broadened Gaussian distributions with an additional term taking into account the relaxation of the singlet state during the measurement~\cite{Barthel09}: 
\begin{equation*}
    N(|S_{21}|)=N_{\mathrm{tot}}[P_S n_S + (1-P_S)n_T]|S_{21}|_{\mathrm{bin}},
\end{equation*}
with 
\begin{equation*}
    n_T=\frac{1}{\sqrt{2\pi}\sigma_T}e^{-\frac{(|S_{21}|-\mu_T)^2}{2\sigma_T^2}}
\end{equation*}
the triplet probability density and 
\begin{multline*}
    n_S=\frac{1}{\sqrt{2\pi}\sigma_S}e^{-\frac{(|S_{21}|-\mu_S)^2}{2\sigma_S^2}} e^\frac{t_{\mathrm{int}}}{T_1} +\\ \frac{1}{\sqrt{2\pi}\sigma_S} \frac{t_{\mathrm{int}}}{T_1}
\int_{\mu_T}^{\mu_S}\frac{1}{\mu_S-\mu_T}e^{-\frac{x-\mu_S}{\mu_S-\mu_T}} e^{\frac{(|S_{21}|-x)^2}{2\sigma_S^2}}dx
\end{multline*}
the singlet probability density. Here, $\mu_T$ ($\mu_S$) is the average triplet (singlet) signal amplitude, $\sigma_T$ ($\sigma_S$) is the standard deviation of the triplet (singlet) peak, $P_S$ is the probability of loading into S(1,1) and $|S_{21}|_{\mathrm{bin}}$ is the bin size. We note that the singlet peak has a slightly larger spread than the triplet peak. This could be explained by the fact that in addition to the measurement noise that broadens the triplet signal, the singlet signal is also prone to effects of charge noise.

We use the following definition of the read-out fidelities: $F_\mathrm{triplet}=1-\int_{|S_{21}|_{\mathrm{th}}}^{\infty}n_{T}d|S_{21}|$ and $F_\mathrm{singlet}=1-\int_{-\infty}^{|S_{21}|_{\mathrm{th}}}n_{S}d|S_{21}|$. The visibility is defined as $V=F_{\mathrm{triplet}}+F_{\mathrm{singlet}}-1$. The maximum visibility for 9 $\mu$s averaging is 96.9\% (see Fig.~3e). The corresponding read-out fidelity for the singlet (triplet) is 97.3\% (99.5\%), with an average read-out fidelity of 98.4\%. We repeat this analysis for various integration times (see Fig.~3f). The average read-out fidelity is above 98\% for $t_{\mathrm{int}}$ greater than 6 $\mu$s.

In conclusion, we have used a high-$Q$ and high-impedance on-chip superconducting resonator to demonstrate single-shot gate-based spin read-out in silicon in a few microseconds. Despite the relatively short $T_1$ in our system, we can still achieve a spin read-out fidelity up to 98.4\% in less than 10 $\mu$s. Extrapolating our results assuming a $T_1$ of 4.5 ms and $t_{\mathrm{int}} = 16$ $\mu$s, we expect a spin read-out fidelity of 99.9$\%$ is possible, well above the fault-tolerance threshold. Further improvements both in the duration and fidelity of spin read-out can be achieved by using quantum-limited amplifiers, such as a JPA or a traveling wave parametric amplifier (TWPA). The  demonstration of single-shot gate-based spin read-out is a crucial step towards read-out in dense spin qubit arrays where it is not possible to integrate electrometers and accompanying reservoirs adjacent to the qubit dots. In contrast, multiple qubits on the inside of an array can be probed using a single resonator coupled to a word or bit line in a cross-bar architecture. In addition, a single feedline can be used for probing multiple resonators using frequency multiplexing.

\section*{Acknowledgments}
We thank T.~F.~Watson, J.~P.~Dehollain, P.~Harvey-Collard, U.~C.~Mendes, B.~Hensen and other members of the spin qubit team at QuTech for useful discussions, L.~Kouwenhoven and his team for access to NbTiN films, and P.~Eendebak and L.~Blom for software support. This research was undertaken thanks in part to funding from the European Research Council (ERC Synergy Quantum Computer Lab), the Netherlands Organisation for Scientific Research (NWO/OCW) as part of the Frontiers of Nanoscience (NanoFront) program, and Intel Corporation.


\begin{thebibliography}{35}%
\makeatletter


\providecommand \@ifxundefined [1]{%
	\@ifx{#1\undefined}
}%
\providecommand \@ifnum [1]{%
	\ifnum #1\expandafter \@firstoftwo
	\else \expandafter \@secondoftwo
	\fi
}%
\providecommand \@ifx [1]{%
	\ifx #1\expandafter \@firstoftwo
	\else \expandafter \@secondoftwo
	\fi
}%
\providecommand \natexlab [1]{#1}%
\providecommand \enquote [1]{``#1''}%
\providecommand \bibnamefont [1]{#1}%
\providecommand \bibfnamefont [1]{#1}%
\providecommand \citenamefont [1]{#1}%
\providecommand \href@noop [0]{\@secondoftwo}%
\providecommand \href [0]{\begingroup \@sanitize@url \@href}%
\providecommand \@href[1]{\@@startlink{#1}\@@href}%
\providecommand \@@href[1]{\endgroup#1\@@endlink}%
\providecommand \@sanitize@url [0]{\catcode `\\12\catcode `\$12\catcode
	`\&12\catcode `\#12\catcode `\^12\catcode `\_12\catcode `\%12\relax}%
\providecommand \@@startlink[1]{}%
\providecommand \@@endlink[0]{}%
\providecommand \url [0]{\begingroup\@sanitize@url \@url}%
\providecommand \@url [1]{\endgroup\@href {#1}{\urlprefix }}%
\providecommand \urlprefix [0]{URL }%
\providecommand \Eprint [0]{\href }%
\providecommand \doibase [0]{http://dx.doi.org/}%
\providecommand \selectlanguage [0]{\@gobble}%
\providecommand \bibinfo [0]{\@secondoftwo}%
\providecommand \bibfield [0]{\@secondoftwo}%
\providecommand \translation [1]{[#1]}%
\providecommand \BibitemOpen [0]{}%
\providecommand \bibitemStop [0]{}%
\providecommand \bibitemNoStop [0]{.\EOS\space}%
\providecommand \EOS [0]{\spacefactor3000\relax}%
\providecommand \BibitemShut [1]{\csname bibitem#1\endcsname}%
\let\auto@bib@innerbib\@empty

\bibliographystyle{apsrev}

\bibitem{Zwanenburg13} F. A. Zwanenburg, A. S. Dzurak, Andrea Morello, M. Y. Simmons, L. C. L. Hollenberg, G. Klimeck, S. Rogge, S. N. Coppersmith, M. A. Eriksson, \textit{Rev. Mod. Phys.} \textbf{85}, 961 (2013).

\bibitem{Ono02} K. Ono, D. G. Austing, Y. Tokura, S. Tarucha, \textit{Science} \textbf{297}, 5585 (2002).

\bibitem{Elzerman04} J. M. Elzerman, R. Hanson, L. H. Willems van Beveren, B. Witkamp, L. M. K. Vandersypen, L. P. Kouwenhoven, \textit{Nature} \textbf{430}, 431 (2004).

\bibitem{Hanson07} R. Hanson, L. P. Kouwenhoven, J. R. Petta, S. Tarucha, L. M. K. Vandersypen, \textit{Rev. Mod. Phys.} \textbf{79}, 1217 (2007).

\bibitem{Field93} M. Field, C. G. Smith, M. Pepper, D. A. Ritchie, J. E. F. Frost, G. A. C. Jones, and D. G. Hasko, \textit{Phys. Rev. Lett.} \textbf{70}, 1311 (1993).

\bibitem{Barthel10} C. Barthel, M. Kjærgaard, J. Medford, M. Stopa, C. M. Marcus, M. P. Hanson, and A. C. Gossard, \textit{Phys. Rev. B} \textbf{81}, 161308 (2010).

\bibitem{Barthel09} C. Barthel, D. J. Reilly, C. M. Marcus, M. P. Hanson, and A. C. Gossard, \textit{Phys. Rev. Lett.} \textbf{103}, 160503 (2009).

\bibitem{Vandersypen17} L. M. K. Vandersypen, H. Bluhm, J. S. Clarke, A. S. Dzurak, R. Ishihara, A. Morello, D. J. Reilly, L. R. Schreiber and M. Veldhorst, \textit{npj Q. Info.} \textbf{3}, 34 (2017).

\bibitem{Li18} R. Li, L. Petit, D. P. Franke, J. P. Dehollain, J. Helsen, M. Steudtner, N. K. Thomas, Z. R. Yoscovits, K. J. Singh, S. Wehner, L. M. K. Vandersypen, J. S. Clarke, M. Veldhorst, \textit{Sci. Adv.} \textbf{4}, 7 (2018).

\bibitem{Veldhorst17}M. Veldhorst, H. G. J. Eenink, C. H. Yang, A. S. Dzurak, \textit{Nature Comm.} \textbf{8}, 1766 (2017).

\bibitem{Colless13} J. I. Colless, A. C. Mahoney, J. M. Hornibrook, A. C. Doherty, H. Lu, A. C. Gossard, and D. J. Reilly, \textit{Phys. Rev. Lett.} \textbf{110}, 046805 (2013).

\bibitem{Zalba15} M. F. Gonzalez-Zalba, S. Barraud, A. J. Ferguson, A. C. Betz, \textit{Nature Comm.} \textbf{6}, 6084 (2015).

\bibitem{Betz15} A. C. Betz, R. Wacquez, M. Vinet, X. Jehl, A. L. Saraiva, M. Sanquer, A. J. Ferguson, M. F. Gonzalez-Zalba, \textit{Nano Lett.} \textbf{15}, 4622 (2015).

\bibitem{House15} M. G. House, T. Kobayashi, B. Weber, S. J. Hile, T. F. Watson, J. van der Heijden, S. Rogge, M. Y. Simmons, \textit{Nature Comm.} \textbf{6}, 8848 (2015).

\bibitem{Rossi17} A. Rossi, R. Zhao, A. S. Dzurak, M. F. Gonzalez-Zalba, \textit{Appl. Phys. Lett.} \textbf{110}, 212101 (2017).

\bibitem{Ahmed18} I. Ahmed, J. A. Haigh, S. Schaal, S. Barraud, Y. Zhu, C.-M. Lee, Ma. Amado, J. W. A. Robinson, A. Rossi, J. J. L. Morton, M. F. Gonzalez-Zalba, \textit{Phys. Rev. Applied} \textbf{10}, 014018 (2018).

\bibitem{Pakkiam18} P. Pakkiam, A. V. Timofeev, M. G. House, M. R. Hogg, T. Kobayashi, M. Koch, S. Rogge, and M. Y. Simmons, \textit{Phys. Rev. X} \textbf{8}, 041032 (2018).

\bibitem{West18} A. West, B. Hensen, A. Jouan, T. Tanttu, C.H. Yang, A. Rossi, M.F. Gonzalez-Zalba, F.E. Hudson, A. Morello, D.J. Reilly, A.S. Dzurak, arXiv:1809.01864.

\bibitem{Urdampilleta18} M. Urdampilleta, D. J. Niegemann, E. Chanrion, B. Jadot, C. Spence, P.-A. Mortemousque, C. Bäuerle, L. Hutin, B. Bertrand, S. Barraud, R. Maurand, M. Sanquer, X. Jehl, S. De Franceschi, M. Vinet, T. Meunier, arXiv:1809.04584 (2018).

\bibitem{Fowler12} A. G. Fowler, M. Mariantoni, J. M. Martinis, A. N. Cleland, \textit{Phys. Rev. A} \textbf{86}, 032324 (2012).

\bibitem{Samkharadze18} N. Samkharadze, G. Zheng, N. Kalhor, D. Brousse, A. Sammak, U. C. Mendes, A. Blais, G. Scappucci, L. M. K. Vandersypen, \textit{Science} \textbf{359}, 6380 (2018).

\bibitem{Rochette17} S. Rochette, M. Rudolph, A.-M. Roy, M. Curry, G. Ten Eyck, R. Manginell, J. Wendt, T. Pluym, S. M. Carr, D. Ward, M. P. Lilly, M. S. Carroll, M. Pioro-Ladrière, arXiv:1707.03895.

\bibitem{Samkharadze16} N. Samkharadze, A. Bruno, P. Scarlino, G. Zheng, D. P. DiVincenzo, L. DiCarlo, and L. M. K. Vandersypen, \textit{Phys. Rev. Applied} \textbf{5}, 044004 (2016).

\bibitem{Cottet11} A. Cottet, C. Mora, and T. Kontos, \textit{Phys. Rev. B} \textbf{83}, 121311 (2011).

\bibitem{Petersson12} K. D. Petersson, L. W. McFaul, M. D. Schroer, M. Jung, J. M. Taylor, A. A. Houck, J. R. Petta, \textit{Nature} \textbf{490}, 380 (2012).

\bibitem{Frey12} T. Frey, P. J. Leek, M. Beck, A. Blais, T. Ihn, K. Ensslin, and A. Wallraff, \textit{Phys. Rev. Lett.} \textbf{108}, 046807 (2012).

\bibitem{Schroer12} M. D. Schroer, M. Jung, K. D. Petersson, and J. R. Petta, \textit{Phys. Rev. Lett.} \textbf{109}, 166804 (2012).

\bibitem{Chorley12} S. J. Chorley, J. Wabnig, Z. V. Penfold-Fitch, K. D. Petersson, J. Frake, C. G. Smith, and M. R. Buitelaar, \textit{Phys. Rev. Lett.} \textbf{108}, 036802 (2012).

\bibitem{Stehlik15} J. Stehlik, Y.-Y. Liu, C. M. Quintana, C. Eichler, T. R. Hartke, and J. R. Petta, \textit{Phys. Rev. Applied} \textbf{4}, 014018 (2015).

\bibitem{Mi17} X. Mi, J. V. Cady, D. M. Zajac, P. W. Deelman, and J. R. Petta, \textit{Science} \textbf{335}, 156 (2017).

\bibitem{Stockklauser17} A. Stockklauser, P. Scarlino, J. V. Koski, S. Gasparainetti, C. K. Andersen, C. Reichl, W. Wegscheider, T. Ihn, K. Ennslin, and A. Wallraff, \textit{Phys. Rev. X} \textbf{7}, 011030 (2017).

\bibitem{Bruhat18} L. E. Bruhat, T. Cubaynes, J.J. Viennot, M. C. Dartiailh, M.M. Desjardins, A. Cottet, T. Kontos, \textit{Phys. Rev. B} \textbf{98} 155313 (2018).

\bibitem{Schuster05} D. I. Schuster, A. Wallraff, A. Blais, L. Frunzio, R.-S. Huang, J. Majer, S. M. Girvin, and R. J. Schoelkopf, \textit{Phys. Rev. Lett.} \textbf{94}, 123602 (2005).

\bibitem{Mizuta17} R. Mizuta, R. M. Otxoa, A. C. Betz, M. F. Gonzalez-Zalba, \textit{Phys. Rev. B} \textbf{95}, 045414 (2017).

\bibitem{Khalil12} M. S. Khalil, M. J. A. Stoutimore, F. C. Wellstood, K. D. Osborn, \textit{J. App. Phys.} \textbf{111}, 054510 (2012).


\end{thebibliography}
\end{document}